\documentstyle[prl,aps,epsfig]{revtex}
\begin{document}

\draft
\flushbottom
\twocolumn[\hsize\textwidth\columnwidth\hsize
\csname@twocolumnfalse\endcsname
\title {Anderson-Mott Transition Driven by Spin Disorder:\\
Spin Glass Transition and Magnetotransport in Amorphous GdSi}

\author{Pinaki Majumdar and Sanjeev Kumar}

\address{ Harish-Chandra  Research Institute,\\
 Chhatnag Road, Jhusi, Allahabad 211 019, India }

\date{Apr 23,  2003}

\maketitle
\tightenlines
\widetext
\advance\leftskip by 57pt
\advance\rightskip by 57pt
\begin{abstract}

A  zero temperature Anderson-Mott transition  driven by spin 
disorder can be `tuned' by an applied  magnetic field to 
achieve  colossal magnetoconductance. Usually this is not possible
since spin disorder by itself cannot localise a high density electron
system. However, the presence of strong structural disorder can realise
this situation, self consistently generating a disordered magnetic
ground state.  We explore such a model, constructed to understand amorphous 
GdSi, and highlight the emergence of a spin glass phase, Anderson-Mott 
signatures in transport and tunneling spectra, and unusual magneto-optical
conductivity.  We solve a disordered strong coupling fermion-spin-lattice 
problem essentially exactly on finite systems, and  account for all the 
qualitative features observed in magnetism, transport, and the optical 
spectra in this system.

\

\

\end{abstract}

]

\narrowtext
\tightenlines

\noindent
The `Anderson-Mott' insulator-metal transition (IMT) in 
disordered interacting systems \cite{and-loc-revs,mott-and-refs}
and the physics of `colossal magnetoresistance' (CMR) in,
for example, the manganese oxides \cite{mang-revs}
are topics of great current interest. 
Even though the ``ultimate'' example of CMR would be a 
magnetic field driven, zero temperature,
insulator-metal transition, such a scenario had not been
realised experimentally till recently.  
Most observations of CMR are at finite temperature, across a 
 ferromagnet
to paramagnet transition \cite{mang-revs,mag-sem-revs,oldcmr}.
{\it Canonical `Anderson-Mott' systems do not show large
magnetoconductance (MC) and the standard CMR systems do not
involve localisation physics.}
These two fields
of research
have evolved independently. 

Experiments \cite{hellman1,hellman2,hellman3,hellman4,hellman5} 
on amorphous $a$-GdSi reveal that the presence of doped  magnetic 
moments in a strongly disordered 
system can combine features of the  standard
doping driven IMT in amorphous systems \cite{nbsi-ref} with the
physics of field driven IMT and CMR.
The magnetic ground state in such a system is a spin glass.
There are  distinct signatures of electron correlation in the
conductivity and tunneling spectra, and huge 
transfer of optical spectral weight on application of
a magnetic field. 
The experimental observations, discussed below, 
cannot be understood within the standard 
scenarios \cite{and-loc-revs,mott-and-refs,mang-revs,mag-sem-revs,oldcmr}
developed for IMT and CMR and require an independent
and comprehensive framework.
Our main achievement in this paper is $(i)$~to provide the
first 
understanding of the unique properties
of this system, and $(ii)$~demonstrate a many body technique 
that allows controlled approximations in a strongly disordered
interacting system.

The measurements  have been made 
on $a$-GdSi and simultaneously on the  non-magnetic analog
$a$-YSi. 
$(i)$~Both Y$_x$Si$_{1-x}$  and Gd$_x$Si$_{1-x}$ show an  
IMT 
\cite{hellman1} as the doping, $x$, is increased 
beyond a critical value, $x_c$. 
The critical doping  $ x_c \sim 14 \%$ in YSi and
$x_c \sim 15 \%$ in GdSi. 
$(ii)$~For $x \lesssim x_c$ in GdSi, a magnetic field, $h_c(x)$, 
 can actually drive
an insulating sample metallic. 
YSi samples show weak  {\it positive} magnetoresistance.
$(iii)$~The density of states (DOS)
 at the Fermi level, $N(0)$, in GdSi, probed 
through  tunneling conductance measurement \cite{hellman2},
grows as  $( h - h_c )^2$ across the IMT, while 
$\sigma_{dc}$ increases as $(h - h_c)$.  
$(iv)$ The optical conductivity in GdSi shows large transfer of spectral 
weight \cite{hellman3}  to low frequency from $\omega \gtrsim 0.1$~eV on 
application of
a field of a few Tesla. Transfer of spectral weight also 
occurs on raising
temperature, and this is seen in both GdSi and YSi.
$(v)$~The low field a.c susceptibility in GdSi 
reveals \cite{hellman4}  that the magnetic
degrees of freedom freeze into a spin glass state 
at low
temperature.
The freezing temperature, $T_f$, increases 
from $ \sim 1$~K at
$x= 0.04$ to $T_f \sim 6$~K at $x \sim 0.20$.
The Y doped samples are {\it  diamagnetic}.
$(vi)$ The `effective moment' inferred \cite{hellman4} from $\chi(T)$
differs from the expected value  for Gd, and the high temperature
magnetic specific heat \cite{hellman5} per doped Gd
is almost $50 \%$ larger than $log(2S + 1)$.

Observation $(i)$ above is standard in disordered systems, 
$(ii) - (iv)$ would be expected in CMR materials, and $(v)-(vi)$ seem
to be unique to the combination.

A `first principles' model for amorphous  GdSi will have to consider 
an underlying `random' structure in which a fraction $x$ of the sites are
occupied by Gd atoms and $(1-x)$ by Si atoms. The Gd and Si atoms have 
different orbital structure so a complicated set of inter-orbital,
intersite hopping possibilities need to be considered. 
We try to retain the essential features in the following, simpler,
one band model:

\begin{eqnarray}
H && = -t \sum_{\langle ij \rangle, \sigma } 
c^{\dagger}_{i \sigma}c_{j \sigma}  
+ \sum_{i\sigma} (\epsilon_i - \mu) n_{i \sigma} 
-J'\sum_{\nu} {\bf \sigma}_{\nu}.{\bf S}_{\nu} 
\cr
&&~~~~~~~~~
 -\lambda \sum_{\nu}  n_{\nu}  x_{\nu} 
+ {1 \over 2}K\sum_{\nu}  x_{\nu}^2 
+  H_{Coul}
\end{eqnarray}

We use a  
tight binding model with uniform hopping 
$t$, and an on site potential  $\epsilon_i$ 
uniformly distributed between $\pm \Delta/2$.   
The sites labelled `$\nu$' are a fraction $x$ of the lattice 
corresponding to the dopant (Y/Gd) 
 locations. 
The electron-spin coupling is $J'$, ${\bf S}_i$ are
the $S = 7/2$ Gd spins,
and $\sigma_{i\mu} = \sum_{\alpha \beta}
c^{\dagger}_{i \alpha} \tau_{\mu}^{\alpha \beta}
c_{i \beta}$, where $\tau_{\mu}$ are the Pauli matrices.
The $x_i$ are local displacement variables (bond 
distortions \cite{pwa-amorph}) 
coupled to the electron density via $\lambda$, and the structural stiffness
is $K$. 
$H_{Coul}$ would include Hubbard and long range Coulomb interactions.

The width of the impurity level distribution $(\Delta)$ 
in YSi/GdSi 
has been estimated \cite{hellman3} to be  $\sim 200$~meV,
 and the `polaron
binding energy' $g = \lambda^2/K \sim 30$~meV. 
The existence of 
lattice polaron effects in amorphous
semiconductors had been argued early on by Anderson \cite{pwa-amorph},
and has been revived now \cite{hellman3} in the context of doped $a$-Si.
We think the diamagnetism in YSi confirms  bipolaronic 
effects, but these 
lattice effects are probably not very important in GdSi.
There is no simple estimate of the `effective hopping amplitude' 
to be used in a single band approximation. 
However, calculations on the Anderson model indicate \cite{economu}
that we need $\Delta/t \approx 14$ to localise $10 $\% of the 
electronic states in the band. This suggests a rather small effective 
hopping amplitude $ \sim 200$~K, if $\Delta$  is $200$~meV. 
 The electron-spin coupling $J'S$ (called $J'$ from now on), arising out of the
$d-f$ coupling in Gd, is large. It is estimated to be $\sim 0.9$~eV from
photoemission measurements \cite{maiti-pes} on Gd, but would be somewhat 
smaller in the effective one band description that we are using.
Although 
the parameter values have some uncertainty it is clear that $\Delta,
J' \gg t$. We use 
$\Delta/t = 11$, $J'/t =4$, $g/t = 0.5$, roughly consistent
with the experimental estimates. 
The electron-phonon and electron-spin coupling are operative only 
at the
dopant sites. We  distribute 
the `impurities'
(Y or Gd in the Si host) into $Nx$ sites with the lowest potential
in any given realisation $\{ \epsilon_i \}$.
This ensures that at low dopant concentration, the electrons are
trapped near the dopant sites.
We  measure all energies in units of $t$, and   
finally assume $t \approx 500$~K for comparing our energy scales 
 with the data.

\begin{center}
\begin{figure}
\epsfxsize=6.0cm \epsfysize=5.7cm \epsfbox{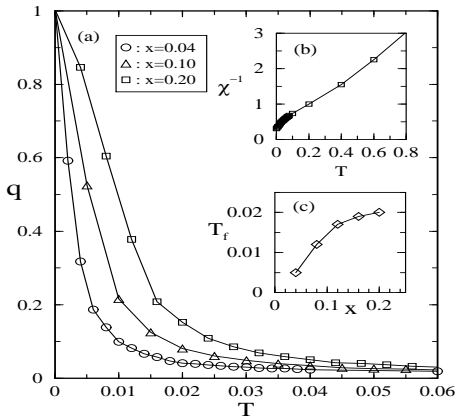}
\caption{$(a).$  Order parameter for freezing, $q(T)$ (see text), 
within  our effective magnetic model. Insets: 
$(b).$ The inverse susceptibility
${\chi}^{-1}(T)$ over a large temperature range. 
$(c).$ Freezing temperature $T_f(x)$. 
Simulation on $10^3$ lattices.
}
\end{figure}
\end{center}

\begin{center}
\begin{figure}
\epsfxsize=7.cm \epsfysize=5.3cm \epsfbox{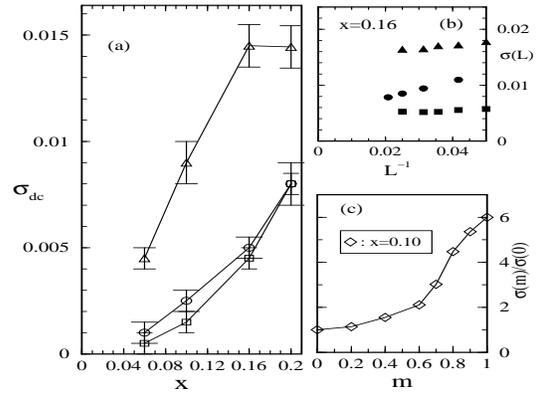}
\caption{$(a)$: D.C conductivity, at $T=0$. 
Model for: YSi (circles), 
GdSi in the spin glass phase (squares), and 
GdSi in the fully polarised
phase (triangles). 
 $(b)$: $L \rightarrow \infty$ extrapolation to construct 
$\sigma_{dc}$ (see text). $(c)$: GdSi: dependence of $\sigma_{dc}$, at $x=0.10$
and $T=0$, on   
magnetisation.}
\end{figure}
\end{center}

Our principal results using the Hamiltonian above are the
following  $(i)$~the magnetic ground state in the 
low doping region is 
a spin glass with $T_f$ having overall 
scale $ \sim t^2/(J'+ \Delta)$ and following the 
experimental doping dependence (Fig.1),
$(ii)$~there is a metal-insulator transition for both
GdSi and YSi with decreasing $x$, with $x_{Gd}^c \gtrsim x_Y^c$, 
and a field driven IMT and CMR for GdSi (Fig.2), 
$(iii)$~there is 
large transfer of spectral weight from high to low frequency (Fig.4)
in $\sigma(\omega)$,  driven by an applied magnetic field in GdSi.
We explain our scheme of calculation next and then discuss these
results in detail.

Since the Hamiltonian involves $\Delta/t, J'/t  \gg 1 $ 
and $g \sim {\cal O}(t) $ none of
these couplings can be handled perturbatively. 
To study the properties of this model within a 
controlled approximation we use
a finite size combination of Monte-Carlo and exact 
diagonalisation \cite{mc-ed-ref} (MC+ED). 
This approach exactly handles the  
strong disorder,  but treats 
the spin and lattice variables as `classical'. 
Since we have  $2S \gg 1$, 
the `classical' spin limit should be 
a reasonable starting point.
At strong disorder, the leading effect of phonons
should also be accessible classically.

If we ignore $H_{coul}$ to start with,  $H$ represents 
non-interacting fermions coupled to classical
variables $ {\bf S}_i$ and $x_i$, in addition to the random
potential $\epsilon_i$. The $\epsilon_i$ are `quenched'
variables   
while the spin and lattice degrees of freedom
are `annealed', with the distribution 
$P\{ x, S \}   
= Z^{-1} Tr e^{-\beta H} $
where  $Z =   \int {\cal D}S {\cal D}x Tr e^{-\beta H}$
is the full partition function for a specific realisation of 
$\{ \epsilon_i \}$.

The `exact' MC+ED allows only small system sizes,
${\cal O}(100)$ sites, so  
the key step is to  
construct an  approximate `effective Hamiltonian'
for the lattice and spin variables.
Once the magnetic and phonon problem are
self consistently solved, the $T=0$ electron problem can be solved
in the classical ground state $\{{\bf S}_i,
x_i \}_0$, which  itself
depends  on  $\{ \epsilon_i \}$, finally averaging over disorder.

Formally the magnetic 
effective Hamiltonian is  
$ H_{eff} \{S\} =  -{1 \over \beta}
~log \int {\cal D}x Tr e^{-\beta H}  $.
The magnetic problem involves $J'/t \gg 1$,  a dilute 
system (the spins occupy only a fraction $x$ of sites), and strong
disorder in the electron system.
There is no perturbative expansion possible in $J'$
but the large $J'$ and $x \ll 1$ allows a simplification.
In this  limit  
the doped carriers are essentially localised at the magnetic sites, with
the electron density falling off exponentially away from the sites.
This generates a pairwise {\it antiferromagnetic 
coupling}, 
for $R_{ij} > 1$, 
with $J_{ij} \sim (t^2/J')e^{-R_{ij}/\lambda(J')}$, with
$\lambda \propto 1/J'$. 
For neighbouring sites there is a 
ferromagnetic coupling $\sim {\cal O}(t)$.

\begin{center}
\begin{figure}
\epsfxsize=6.5cm \epsfysize=6.0cm \epsfbox{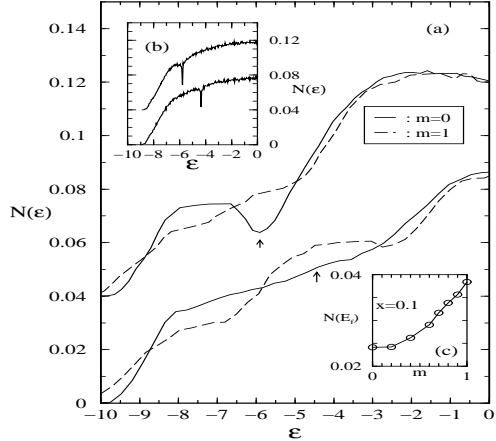}
\caption{$(a).$ Density of states in the model for 
GdSi for two densities,
$x=0.1$ and $x=0.2$, 
for random spins $(m=0)$ and polarised spins
$(m=1)$. The curves are vertically shifted by $0.04$.
Fermi energy marked by arrows.
 $(b).$ DOS in model for 
YSi, at $x=0.1$ (above) and $x=0.2$ (below).
$(c)$. DOS at $\epsilon_F$ in the model for 
GdSi, at $x=0.1$, with
increasing magnetisation. }
\end{figure}
\end{center}

In the disordered system, for a given separation 
$R_{ij}$, the bonds have a distribution.  We construct 
the distributions $P(J, R_{ij})$ and study the model:
$
H_{eff}\{{\bf S} \} =  \sum_{ij}J_{ij} {\bf S}_i.{\bf S}_j.
$
For a pair of moments located at ${\bf R}_i$ and ${\bf R}_j$ 
the $J_{ij}$ is picked from $P(J, R_{ij} )$.
This ignores correlations between bonds in a specific
$\{ \epsilon_i \}$ realisation. 
We simulate the model for different dilutions,  
compute the  order parameter for freezing,
$q(T) = (xN)^{-1} \sum_i
\vert \langle {\bf S}_i \rangle_T \vert $, and check that
the structure factor has no peaks at any wavevector ${\bf Q}$.
Fig.1 shows $q(T)$, alongwith $\chi(T)$ at $x=0.2$, and the
freezing temperature  \cite{sg-heis-ref} $T_f(x)$.
At $x=0.2$ our $T_f \sim 10$K, while experimentally $T_f
\sim 6$K.
We have done the exact MC+ED simulation for $4^3$ systems,
with the same $J', \Delta$ and  $x \sim 0.1$ and
verified \cite{longpap}
that the system freezes into a spin glass, with $T_f$
within $10 \%$ of our result here.

Having established  the
existence of a glassy state 
for the $\{ {\bf S}_i \}$
we will simplify the remaining electron-phonon problem by 
assuming the spins to be frozen in an uncorrelated random manner.

The  effective  
Hamiltonian for phonons is 
$H_{eff} \{x\} =  -{1 \over \beta}
~log \int {\cal D}S Tr e^{-\beta H} $.
At moderate $g$ and strong disorder there would be `frozen' bond 
distortions in the ground state, and we cannot expand about the
$x_i=0$ state.  
To incorporate this effect we use the lowest order self-consistent
expansion, {\it i.e},
$
H_{eff} \{x\} \approx  {1 \over 2} K \sum_i x_i^2  + \sum_i a_i x_i  
$
with  $a_i = -\lambda {\bar n_i}$, where ${\bar n_i} 
= \langle n_i \rangle $, computed in 
the electronic ground state.
The minimum  of $H_{eff}$, {\it i.e} the lattice
distortion  in the ground state, 
corresponds to ${\bar x_i} = (\lambda/K) {\bar n_i}$.
The $T=0$ problem now corresponds to electrons in
the background of structural disorder, $\{ \epsilon_i \}$,
coupled to randomly oriented spins with coupling $J'$, and 
density coupled to a phonon field $x_i = (\lambda/K){\bar n_i}$,
{\it i.e}, 
\begin{equation}
H_{eff}^{el} = 
H_{kin}
+ \sum_{i} \epsilon_i n_{i} 
-J'\sum_{\nu} {\bf \sigma}_{\nu}.{\bf S}_{\nu} 
 -g\sum_{\nu} {\bar n_{\nu}} n_{\nu}
\end{equation}
We solve this problem through iterative ED. 
The transport and spectral properties of YSi correspond to 
$g=0.5$ and  $J'=0$, while for GdSi  $g=0.5$ and $J'=4$.
The ED  is done for a sequence of
sizes $6 \times 6 \times L$, with $L=24, 32, 40, 48$. 
Due to the finite size gaps the d.c. conductivity cannot be 
directly computed on finite systems.
We use the Kubo-Greenwood
formula 
to compute the integrated optical
spectral weight ${\sigma_{int}(\Delta \omega)}
= \int_0^{\Delta \omega} \sigma(\omega) d\omega$,
disorder average, and invert to obtain the optical
conductivity $\sigma(\omega)$. 
The extent of averaging varies from $400-100$ realisations, 
decreasing with increasing $L$. 
We track 
$\sigma(\omega_{ref};L)$  with $\omega_{ref} \propto L^{-1}$,
setting  
$\omega_{ref}=0.08$ at $L=32$, and use  
$\sigma_{dc} =
{lim}_{L \rightarrow \infty}  \sigma(\omega_{ref},L)$.  
Fig.2$(a)$ shows this `dc' conductivity appropriate 
to YSi and GdSi. 
Our $x_c$ are smaller than the experimental values and we have
not fine tuned parameters to match the data.
The  $\sigma$ are in units of $\pi e^2/{\hbar a_0}$ and the
typical values shown in Fig.2$(a)$ are $\sim 0.01$. 
For $a_0 \sim 2 \AA$,  $\sigma_{dc} \sim 400$ $(\Omega cm)^{-1}$,
roughly as in experiments \cite{hellman1}.
Fig.2$(b)$ shows the $L$ dependence of $\sigma(\omega_{ref} ; L)$,
while Fig.2$(c)$ shows the `magnetoconductance'.
We have checked that the conductivity in a `spin glass' backgound 
shows the same trend as for random spins and the numbers match 
within $\sim 20 \%$.

\begin{center}
\begin{figure}
\epsfxsize=6.5cm \epsfysize=6.1cm \epsfbox{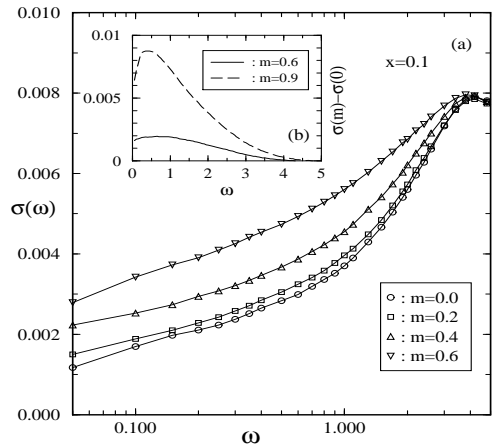}
\caption{Variation of optical conductivity in the model for 
GdSi, at $T=0$, with degree
of magnetisation. Note the log scale in frequency.
Inset: magneto-optical conductivity.}
\end{figure}
\end{center}

To our knowledge there are no standard results
on  electron systems combining strong structural 
disorder {\it and} strong coupling
to dilute magnetic moments.
In the well studied opposite limit,  $\Delta/t \gg1, 
J'/t \ll 1$, 
spin flip scattering actually
{\it weakens} \cite{and-loc-revs}
 Anderson localisation, so large $J'/t$ is crucial
to enhanced localisation in GdSi.
We have cross checked the trends in $ \sigma_{dc}$
by computing the averaged Greens function 
$G^{\sigma \sigma'} ({\bf r} -{\bf r}')$
 at large $\Delta$, with increasing $J'$. 
Using $Tr G = \sum_{\sigma} G^{\sigma \sigma} ({\bf r} -{\bf r}')$
as the indicator of `delocalisation', we find that  
at $\vert {\bf r} -{\bf r}' \vert/a_0 = 6$,
$Tr G$ {\it grows} with $J'$ upto $J' \sim 0.5$
and then falls rapidly \cite{longpap}. 
It recovers quickly as the spins are polarised by an applied
field, tracking the change in conductivity. 

The DOS in GdSi, at low $x$, has a broad minimum 
 at $\epsilon_F$, Fig.3$(a)$, since
$J'$ pulls down states to lower  energy. This minimum is {\it not}
related to the Altshuler-Aronov `correlation gap' 
which would
be a sharper feature 
\cite{mott-and-refs} 
near $\epsilon_F$ (with effects of
$H_{coul}$  included). 
In YSi, the effect of phonons on the disordered background 
shows up as a sharp dip \cite{and-loc-revs} in the DOS, Fig.3$(b)$,
since it generates a  short range attractive 
interaction with $U_{eff} = -\lambda^2/K$.

We mimic the effect of finite magnetisation $(m)$, in GdSi,
by using a spin distribution with
finite $\langle S_z \rangle$. Finite $m$ leads to significant 
redistribution of weight in the DOS, due to the large $J'$,
which should be visible
in photoemission measurements. The conjunction of increased
mobility, and increased DOS near $\epsilon_F$,
Fig.3$(c)$,
leads to the large changes observed  in $\sigma(\omega)$, Fig.4.
The `outer scale' in $\sigma(\omega)$
is $\sim 5t \approx 0.25$~eV, as in the 
data \cite{hellman3}.

There are certain experimental 
features for which the `Mott' aspect is essential.
These are
principally the $\sqrt {T}$ dependence in  $\sigma_{dc}(T)$, the $\sqrt
\omega$ correlation gap, the $T$ driven spectral weight transfer in
$\sigma(\omega)$, and the excess magnetic $C_V$.
Most of these are generic correlation effects, well known in
other amorphous systems \cite{and-loc-revs,nbsi-ref}, and unrelated to the 
magnetic character. 
 
Let us
re-emphasize the uniqueness of the system we study.
Disorder, electron-spin coupling and electron-phonon 
interactions  are features common, in some form, to $a$-GdSi,
Anderson-Mott systems (NbSi, say) and the CMR manganites. 
The crucial differences are:
$(i)$~GdSi is a {\it strongly disordered
`dilute' magnetic system, with strong electron-spin coupling.} These features
are essential to the spin glass behaviour and the consequent IMT and CMR.
Electron-phonon interactions, even if present, are not crucial to the physics.
$(ii)$~Anderson-Mott systems are also {\it strongly disordered, 
but nominally non magnetic}. There are no 
remarkable magnetic field effects and the physics is controlled by 
disorder and electron correlations. 
$(iii)$~Most CMR manganites are reasonable
metals at low temperature, indicating {\it weak~ intrinsic disorder}. 
They have {\it strong electron-spin coupling} on a {\it periodic Mn lattice},
which, in contrast to `dilution', promotes double exchange {\it ferromagnetism}. 
Electron-phonon (Jahn-Teller) interactions are important in these 
systems. 
The finite temperature IMT and CMR are related to multi-phase 
coexistence \cite{mang-revs} and not 
an Anderson transition. GdSi differs also 
from the  diluted magnetic
semiconductors in that the spin polaron concept 
\cite{oldcmr,hellman1} is not tenable
in this high electron density system, due to 
strongly overlapping wavefunctions.

In conclusion, this is the first explanation of insulator-metal transition, 
CMR, spin glass freezing and optical properties of $a$-GdSi, bridging the
gap between Anderson-Mott transition and CMR systems.
Our results are based on an exact finite size calculation, 
handling strong disorder and interactions.
The effect of Coulomb interactions is understood only qualitatively at the
moment, and their inclusion   
would be the next step.

P.M would like to thank T. V. Ramakrishnan, Chandan Dasgupta and
H. R. Krishnamurthy  for discussions. 
We acknowledge use of the Beowulf
cluster at H.R.I.


\begin{thebibliography}{99}

\bibitem{and-loc-revs} P. A. Lee and T. V. Ramakrishnan, Rev. Mod. Phys.
{\bf 57}, 287 (1985).

\bibitem{mott-and-refs} B. L. Altshuler and A. G. Aronov
in {\it Electron-Electron Interactions in Disordered Systems},
ed. A. L. Efros and M. Pollak, North Holland (1985), also
D. Belitz and T. R. Kirkpatrick, Rev. Mod. Phys.
{\bf 66}, 261 (1994).

\bibitem{mang-revs} A. P. Ramirez, J. Phys. Condens Matter {\bf 9}, 8171 (1997), 
E. Dagotto {\it et al.}, : Phys Rep {\bf 344}, 1 (2001),
 {\it Colossal Magnetoresistive Oxides}, Ed. Y. Tokura, Gordon \& Breach (2000)

\bibitem{mag-sem-revs} 
E. L. Nagaev, {\it Physics of Magnetic Semiconductors},
Mir Publishers, Moscow (1983),
also {\it Magnetic Semiconductors} Vol. 25 of  
{\it Semiconductors and Semimetals},
Ed.  Willardson and Beer, Academic Press (1988). 

\bibitem{oldcmr}  
K. Kasuya and A. Yanase, Rev. Mod. Phys. {\bf 40},
684 (1968),
S. von Molnar {\it et al.}, 
Phys. Rev. Lett. {\bf 51}, 706 (1983), 
H. Ohno {\it et al.}, 
 Phys. Rev. Lett. {\bf 68}, 2664 (1992). 

\bibitem{hellman1} F. Hellman, {\it et al.},
Phys. Rev. Lett. {\bf 77}, 4652 (1996).

\bibitem{hellman2} W. Teizer, {\it et al.}
Phys. Rev. Lett. {\bf 85}, 848  (2000).

\bibitem{hellman3} D. N. Basov {\it et al.}, preprint, cond-mat 
0104245.

\bibitem{hellman4} F. Hellman, {\it et al.} 
Phys. Rev. Lett. {\bf 84}, 5411  (2000).

\bibitem{hellman5} B. L. Zink, {\it et al.}, 
Phys. Rev. Lett. {\bf 83}, 2266 (1999).

\bibitem{nbsi-ref} G. Hertel {\it et al.} 
Phys. Rev. Lett. {\bf 50}, 743 (1983). 

\bibitem{pwa-amorph} P. W. Anderson, Phys. Rev. Lett. {\bf 34},
953 (1975)

\bibitem{economu} E. N. Economou {\it et al.}, Phys. Rev. {\bf B 30},
1686 (1984)

\bibitem{maiti-pes} K. Maiti {\it et al.}, Phys. Rev. Lett.
{\bf 88}, 167205-1 (2002).

\bibitem{mc-ed-ref} E. Dagotto {\it et al.}, 
Phys. Rev. {\bf B 58}, 6414 (1998),  
M. J. Calderon and L. Brey, Phys. Rev. {\bf B 58}, 3286 (1998).


\bibitem{sg-heis-ref} 
It is believed that the lower critical dimension for Heisenberg  
spin glasses is three, so the finite $T_f$ seen in small size 
simulation could be an artifact. However, even weak anisotropy,
presumably dipolar couplings in GdSi, can stabilise a finite
$T_f$, and the {\it doping dependence} would be essentially as
seen here. References:
A. J. Bray {\it et al.},  
Phys. Rev. Lett. {\bf 56}, 2641 (1986), 
A. Chakrabarti and C. Dasgupta, Phys. Rev. Lett.
{\bf 56}, 1404 (1986). 

\bibitem{longpap} Sanjeev Kumar and Pinaki Majumdar,  to be
published.




\end{thebibliography}
\end{document}